\documentstyle[preprint,prb,aps]{revtex}

\begin{document}
\draft

\title{Enhanced Charge and Spin Currents in the One-Dimensional
Disordered Mesoscopic Hubbard Ring}

\author{Rudolf A.\ R\"{o}mer}
\address{Condensed Matter Theory Unit,
Jawaharlal Nehru Centre for Advanced Scientific Research,
Indian Institute of Science Campus,
Bangalore, 560 012,
India}

\author{Alexander Punnoose}
\address{Department of Physics,
         Indian Institute of Science,
         Bangalore, 560 012, India}

\date{Version: May 15, 1995; printed \today}
\maketitle

\begin{abstract}
We consider a one-dimensional mesoscopic Hubbard ring with and
without disorder and compute charge and spin stiffness as a measure
of the permanent currents.
For finite disorder we identify critical disorder strength beyond which
the charge currents in a system with repulsive interactions are {\em
larger} than those for a free system.
The spin currents in the disordered repulsive Hubbard model are enhanced
only for small $U$, where the magnetic state of the system corresponds to
a charge density wave pinned to the impurities.
For large $U$, the state of the system corresponds to localized isolated
spins and the spin currents are found to be suppressed.
For the attractive Hubbard model we find that the charge
currents are always suppressed compared to the free system at all
length scales.
\end{abstract}

\pacs{72.15Rn, 73.20Mf}

\narrowtext

%
%

\section{Introduction}

Over the last few years various experiments have measured the
magnetic response of small, moderately disordered ensembles of
quasi one-dimensional (1D) rings.\cite{levy,chandra,mailly}
These experiments confirm the existence of a persistent
current which had been predicted already much earlier.\cite{imry}
However, the experimental value for the persistent current is
two to three orders of magnitude {\em larger} than the theoretical
predictions based on calculations in a disordered but non interacting
electron gas.\cite{boris}
It is thus commonly believed that the discrepancy could probably be
resolved by accurately including the electron-electron interactions
in these calculations.

Studying the full interacting electron problem with disorder is in
general difficult.
However, in one dimension, powerful analytical tools, such as
Bethe Ansatz\cite{bill,kor} and Bosonization\cite{sol,emery} give us a
handle to
treat at least the on-site part of the interaction exactly and
the effect of disorder can then be studied within a perturbative
renormalization group (RG) approach.
Using these methods, Giamarchi and Shastry\cite{tg} have shown that a
repulsive interaction {\em enhances} the value of the persistent
current in a mesoscopic Hubbard ring.
This result has been confirmed by recent independent numerical\cite{kamal}
and analytical\cite{ramin} studies treating the Hubbard interaction in
first order perturbation theory.

In this paper we study the persistent currents in the 1D mesoscopic
Hubbard ring both with and without disorder along the lines of
Ref.\onlinecite{tg}.
Due to the absence of Galilean invariance,\cite{muller} we note that the
persistent currents exhibit a strong reduction as a function of both
filling and interaction strength already for the {\em clean} ring.\cite{bs}
Thus it is a priori not clear that this initial reduction can be
compensated completely by disorder as to give a net enhancement of the
current.
Therefore, we use the exact bosonization parameters, obtained from the
Bethe Ansatz solution for the clean ring, as initial values for the
solution of the RG equations.
We remark that Ref.\onlinecite{tg} uses only perturbative starting values
for the RG equations and normalizes the current w.r.t.\ non-interacting
values.
Our approach allows us to show that for arbitrarily weak disorder
there is {\em no} enhancement in the currents.
Furthermore, we can identify critical disorder strengths beyond which
the persistent current in the presence of both interactions and disorder
is indeed {\em larger} than the current for the disordered but non
interacting system.

We also study the effect of disorder on the {\em spin} currents
in the Hubbard ring.
We emphasize that even in the presence of Galilean invariance the spin
currents can renormalize non-trivially under interactions.
Physically a non-vanishing spin current implies that the system has
long ranged spin correlations,\cite{bs} and it is thus interesting to
see the effect of disorder on the spin currents.

The plan of the paper is as follows:
In section \ref{sec-stiffcurrent} we define charge and spin stiffness
for a 1D mesoscopic system and relate them to the corresponding
currents.
We briefly discuss the range of validity of these relations.
In section \ref{sec-hub} we first review the Bethe Ansatz (BA) solution
for both the repulsive and the attractive clean Hubbard model.
We then go on to compute the stiffnesses and their conjugate
compressibilities by iterating the BA equations for a mesoscopic
system.
Next, we recall the bosonized description of the Hubbard model
and show how to relate stiffness and compressibility to the new
parameters of the bosonized Hamiltonian.
In section \ref{sec-disorder}, we then include the effects of disorder
into the bosonized model by a RG calculation.
These RG equations are then integrated numerically and we can identify
a crossover from a phase with enhanced spin currents to a phase
in which spin currents are suppressed as the on-site
interaction is increased relative to the initial disorder strength.
We discuss these results in section \ref{sec-conclusion}.

%
%

\section{Persistent currents and stiffness of a 1D mesoscopic system}
\label{sec-stiffcurrent}

For a finite system on a ring of length $L$, the response of the
ground state energy to a finite Aharanov-Bohm (AB) flux is a measure
of the persistent current at $T=0$.
The current is given as
$J =\left. L\ {\partial E_0(\Phi)/{\partial\Phi}}\right|_{\Phi=0}$,
where $E_0(\Phi)$ is the ground state energy of the full interacting
system in the presence of the AB flux ${\Phi}$.
The energy shift of the ground state can be written as
$E_0(\Phi)-E_0(0)\equiv D\Phi^2/L+O(\Phi^4)$,
where
$D \equiv \left. (L/2) { \partial ^2E_0(\Phi)/\partial\Phi^2}\right|_{\Phi
= 0}$
is called the stiffness constant.
$D$ provides an operational definition for the persistent
current for small values of the flux, given as $J=2D\Phi$.
Higher order terms are important when the energy shift is comparable to
the mean energy level spacing in the spectrum of the many-body
system.
In the case of a finite 1D metallic system, the gaps are $O(1/L)$
and so non-quadratic corrections occur when
$\Phi$ is $O(1)$. Level crossings would occur and
perturbation theory would break down for $\Phi$ of order $\pi$.

In a system with up and down spins, the AB flux
for each species can be treated as independent parameters
$\Phi_{\uparrow}$ and $\Phi_{\downarrow}$.
With this freedom, there are two stiffnesses that can be calculated and
hence two currents. In the case when
$\Phi_{\uparrow}= \Phi_{\downarrow}= \Phi_c$,
both species are coupled to the same flux and
the shift in the energy gives the charge stiffness $D_c$.
In a Galilean invariant system this flux couples only to the center of
mass coordinate and hence the persistent {\em charge} current will not
depend on the interaction between the particles.\cite{muller}
However, for lattice models such as the Hubbard model, the center of mass
momentum is conserved only up to the reciprocal lattice vector and we can
observe a non-trivial dependence of the charge current on the interaction.
In the sector of zero magnetization ($S_z=0$), the second choice is to take
$\Phi_{\uparrow}=-\Phi_{\downarrow}= \Phi_s/2$.
In this case the two species are driven in opposite directions
through each other which leads to a non-trivial dependence of the
resulting {\em spin} current on the interactions between
opposite spins even for a Galilean invariant system.
The energy shift in the presence of $\Phi_s$ defines the
spin stiffness $D_s$.
Charge and spin currents are then given as
$J_r = 2D_r\Phi_r$ and
\begin{equation}
D_r = \left. {L\over 2} {
\partial ^2E_0(\Phi_r)\over \partial\Phi_r^2}\right|_{\Phi_r
= 0}, \quad r= c,s.
\label{eq-stiff}
\end{equation}
We note that $D_c$ is related to the d.c.\ part of the conductivity
$\sigma(\omega)=D_{Drude}\delta(\omega)+\sigma_{reg}$ by
$D_{drude} \equiv 2\pi D_c$.\cite{shankar,schulz}

%
%

\section{The 1D Hubbard model without disorder}
\label{sec-hub}

\subsection{The repulsive interaction regime}

\subsubsection{Charge and spin stiffness and the conjugate compressibilities}

The repulsive Hubbard model on a ring of size $L$ threaded by a
spin dependent flux $\Phi_{\sigma}$ is described by the
Hamiltonian
\begin{eqnarray}
H = -t \sum_{\stackrel{\scriptstyle i=1}{\sigma=\uparrow,\downarrow}}^{L}
(e^{i\Phi_{\sigma}/L}c_{i+1,\sigma}^{\dag}
c_{i,\sigma} + h.c) +
U \sum_{i=1}^{L} n_{i,\uparrow}
n_{i,\downarrow}, \quad U > 0,
\label{eq-hubbard}
\end{eqnarray}
where $c_{i,\sigma}^{\dag}$ and $c_{i,\sigma}$ create and
annihilate fermions on site $i$ with spin $\sigma$ and
$n_{i,\sigma}=c_{i,\sigma}^{\dagger}c_{i,\sigma}$ is the
density operator.
We define our energy scales by setting $t=1$ in the following.
Note that this is in fact equivalent to using the Fermi velocity
of the non interacting system $v_f$ as defining energy scales, as
$v_f= 2 t \sin [\pi n/2]$, where $n$ is the density.

The addition of a flux $\Phi_{\sigma}$ is compatible with integrability
and the Bethe Ansatz equations for a chain with
total number of particles $N=N_{\uparrow}+N_{\downarrow}$ and
$M=N_{\downarrow}$ number of down spin
particles are given as\cite{bs}
\begin{mathletters}
\label{eq-barep}
\begin{eqnarray}
L k_n
        &= &2\pi I_n + \Phi_{\uparrow} +2 \sum_{j=1}^{M}
        \arctan{[4(\Lambda_j-\sin{k_n})/U]},\\
2 \sum_{n=1}^{N}\arctan{[4(\Lambda_j-\sin{k_n})/U]}
        &= &2\pi J_j+\Phi_{\downarrow}-\Phi_{\uparrow}}
        +2 \sum_{i\neq j}^{M}\arctan{[2(\Lambda_j-\Lambda_i)/U].
\end{eqnarray}
\end{mathletters}
The quantum numbers $I_n$ and $J_j$ label different states
and statistics. For the fermionic ground state they are taken to be
$\Delta I_n = {M/ 2}\ \mbox{mod}\ 1$ and
$\Delta J_j = {(N-M+1)/ 2}\ \mbox{mod}\ 1$.
The energy of the system in a state corresponding to a solution of
Eq.\ (\ref{eq-barep}) is equal to $E_0(\Phi_{\sigma})=-2 \sum\cos{k_n}$.

In order to study the currents in the ring one should
in principle distinguish between cases of odd or even numbers of particles:
If both $N_{\uparrow}$ and $N_{\downarrow}$ are odd, the number of right and
left
moving particles for each of the species are equal, the
ground state is non-degenerate and the energy is minimum in zero
external flux.
For any other combination, i.e.\ when either $N_{\uparrow}$ or
$N_{\downarrow}$ or both are even, the number of right and left moving
particles are not the same due to the $k=0$ state and hence the energy is
not necessarily a minimum at zero charge or spin flux but rather at
flux values of $\pm \pi$.
For the sake of simplicity, we shall only consider the case of zero
magnetization with $N$ even and $M=N/2$ odd in our numerical
computations.

In Figs.\ \ref{fig-dcdp} and \ref{fig-ds}, we show $D_c$ and
$D_s$ as obtained from an iteration of the BA equations for
a ring of size $L=100$ and various fillings $n=N/L$ and values of
the interaction strength $U$.
Both figures show that there exists a reduction in both stiffnesses
with increasing $U$ for a given filling.
The magnitude of the reduction for $D_c$ is considerably enhanced as
the filling is increased towards the half-filled case.

The other physical quantities that will be needed in the next
section to compute the parameters arising in the bosonized
Hubbard Hamiltonian are the charge compressibility and the spin
susceptibility.
The usual thermodynamic expression for the charge compressibility
is given as
$\chi_c^{-1}=\left.(n^2/ L) \partial ^2 E_0 / \partial n^2 \right|_{\Phi=0}$,
where $n=N/L$ is the density of particles.
For a mesoscopic system, the derivatives should be replaced by
finite differences.
In order to keep the magnetization fixed, the charge compressibility
is then computed by adding both a spin up and a spin down particle to the
ground state configuration; the explicit form of the derivative is
\begin{equation}
\chi_c^{-1}= 2 \times \frac{N^2}{4L} [E_0(N+2)+E_0(N-2)-2E_0(N)].
\label{eq-chic}
\end{equation}
The leading factor of two ensures that a pair of particles is introduced.

Shastry and Sutherland\cite{bs} have shown that the bulk spin
susceptibility $\chi_s$ in the Hubbard model is related to the spin
stiffness by $D_s \chi_s = 1/ 2\pi^2 n^2$ due to a remarkable
property of the BA equations (\ref{eq-barep}).
Thus in the thermodynamic limit we only need to compute either
spin stiffness or susceptibility.
However, for a mesoscopic system, this relation does not hold due to
an avoided level crossing at $\Phi_s=2\pi$.\cite{bpc}
We remark that this
situation is very much as for the Heisenberg-Ising model in the
momentum $\pi$ sector.\cite{res}
Following Ref.\onlinecite{bs}, we then write for the spin
susceptibility
\begin{equation}
\chi_s^{-1}= {N^2\over 2L} \left.[ E_0(N,M=N/2-1)-E_0(N,N/2) \right].
\label{eq-chis}
\end{equation}
Note that in the $M=N/2-1$ sector the ground state has the
quantum numbers $I_n=-(N+1)/2+n$ for $1\leq n \leq N$ and
$J_j=-(N/2+1)/2+j$ for $1 \leq j \leq M$.

The behavior of both $\chi_c$ and $\chi_s$ has been reported previously
in the thermodynamic limit.\cite{shiba,frakor} Our mesoscopic results
are qualitatively the same and differ at the most by $15\%$ from their
thermodynamic values. Thus we refrain from including the corresponding
figures here. However, we emphasize that these small deviations will
be quite important in the following sections.

%
%
\subsubsection{Boson representation of the repulsive Hubbard model}

Away from half filling, the low-energy and large distance
behavior of a one-dimensional fermion system with spin-independent
interactions is described by the Hamiltonian\cite{sol,emery}
\begin{equation}
H = H_{c} + H_{s} +
{2 g_{1\perp}\over (2 \pi \alpha)^2} \int dx \cos{(\sqrt{8} \phi_{s}(x))},
\label{eq-boson}
\end{equation}
where
\begin{equation}
H_{r}={1\over 2\pi}\int dx\left[(v_{r}K_{r})(\pi
\Pi_{r})^2 + \left({v_{r}\over K_{r}} \right)(\partial_x
\phi_{r})^2 \right]
\end{equation}
for $r= c,s$.
This Hamiltonian describes the most general 1D
Hamiltonian with spin conserving interactions,
provided that the proper values for $K_{r}$ and $v_{r}$ are
used.\cite{thierry}
The $c$ and $s$ parts of the Hamiltonian describe the
charge and spin degrees of freedom of the system respectively.
The operator $\Pi_{r}$ is the momentum density conjugate to $\phi_{r}$
and these operators obey Bose-like commutation relations:
$[\Pi_{r}(x),\phi_{r}(x')] = -i \delta_{r , \mu }
\delta(x-x')$.
$\alpha$ is a short-range cutoff parameter of the order of
the lattice constant. The $g_{1\perp}$ term represents scattering
between opposite spins with a momentum transfer close to $2k_f$.
The umklapp scattering transferring two particles from $-k_f$ to $k_f$,
involves a momentum transfer $4k_f=2\pi$ which in the half
filled band ($k_f=\pi/2$) corresponds to a
reciprocal lattice vector. Away from the half-filled case this
term does not contribute and therefore has not been included in
the Hamiltonian (\ref{eq-boson}).
The case $g_{1\perp}=0$ describes independent long-wavelength
oscillations of the charge and spin densities with linear
dispersion relation $\omega_{r}(k)=v_{r}|k|$.
For $g_{1\perp}\neq 0$ the cosine term has to be treated
perturbatively.

For the Hubbard Hamiltonian (\ref{eq-hubbard}) the values for
$K_{c}$ and $v_{c}$ are given by the following relations,
\begin{mathletters}
\label{eq-bosonconstrepc}
\begin{eqnarray}
K_c &=& \pi \frac{n}{2} \sqrt\frac{2 D_c}{\chi_c^{-1}}, \\
v_c &=& {2\over n}\sqrt{{D_c \chi_c^{-1} \over 2}}.
\end{eqnarray}
\end{mathletters}
Here $D_c$ and $\chi_c$ are defined as in Eq.\ (\ref{eq-stiff}) and
(\ref{eq-chic}), respectively, and $n=N/L$ is the density of particles.
For $g_{1\perp}=0$, we can similarly write
\begin{mathletters}
\label{eq-bosonconstreps}
\begin{eqnarray}
K_s &=& 2 \times \pi \frac{n}{2} \sqrt\frac{2 D_s}{\chi_s^{-1}}, \\
v_s &=& 2 \times \frac{2}{n} \sqrt{{D_s \chi_s^{-1}}\over 2},
\end{eqnarray}
\end{mathletters}
with $D_s$ and $\chi_s$ defined as in Eq.\ (\ref{eq-stiff}) and
(\ref{eq-chis}) and the leading factor of $2$ is due to our definition
of the spin flux $\Phi_s$.
However, $D_s$ and $\chi_s$ have been calculated from the exact solution
of the Hubbard model, e.g., {\em including} the $2k_f$ scattering terms
represented by the $g_{1\perp}$ term in the bosonized description of
Eq.\ (\ref{eq-boson}).
Therefore, the identifications of Eq.\ (\ref{eq-bosonconstreps}) are not
exact and only valid perturbatively in $g_{1\perp}$.
Since for the Hubbard model $g_{1\perp}= U$, this means
that Eq.\ (\ref{eq-bosonconstreps}) is valid only for small $U$.

Let us briefly elaborate on this point:
For a macroscopic system $g_{1\perp}$ renormalizes to zero.\cite{g1perp}
The spin stiffness in this limit is given as
$D_s={1\over 4\pi} v_s K_s^*$ where $K_s^*=1$ is the
renormalized value of $K_s$.
However, for a finite system, one generates additional
terms of $O(g_{1\perp})$ coming from the term proportional to
$g_{1\perp}$.
We therefore {\em define} the spin stiffness as $D_s={1\over 4\pi} v_s K_s$
with $v_s$ and $K_s$ given as in Eq.\ (\ref{eq-bosonconstreps}).
When $K_s=1$, we find the simple relation between the spin stiffness and
the spin susceptibility as in last section.
However, for mesoscopic systems
deviation of $K_s$ from its macroscopic fixed point value occurs as a
consequence of the avoided level crossing mentioned earlier.
We find that $K_s$ varies from its non interacting value $K_s=1$
up to $K_s\sim 1.15$ for $U=20$ in a ring of size $L=100$.

%
%
\subsection{The attractive interaction regime}

The attractive Hubbard model is again described by the Hamiltonian
of Eq.\ (\ref{eq-hubbard}), but now $U<0$.
The attraction gives rise to Cooper-like bound pairs of spin up and
spin down particles.
These pairs scatter off each other without diffraction.
Hence for $N$ even and zero magnetization, we can write down effective
BA equations for $M=N/2$ pairs.\cite{bill,br}
Let $\alpha(k_i)=2\arcsin{\left(k_i+i{U\over 4}\right)}$. Then the
BA equations for the pairs are given as,
\begin{equation}
L\ \mbox{Re}\{\alpha(k_i)/2\}=
        2\pi J_i+2\Phi_p +
        \sum_{j=1}^{M}\arctan\left[ {2\over U}(k_j-k_i)\right],
\label{eq-baatt}
\end{equation}
where $\Phi_p=\Phi_{\uparrow}=\Phi_{\downarrow}$ is the flux
acting on the Cooper pair.
The factor of two in front of $\Phi_p$ arises due to the charge of
the Cooper pair.
The quantum numbers for the ground state are
$\Delta J_i = (M+1)/2\ \mbox{mod}\ 1$,
and the corresponding ground state energy is given as
$E_0( \Phi_p )=-4 \sum_{i=1}^{M}
\cos[\mbox{Re} \{\alpha(k_i)/2\}] \cosh[\mbox{Im} \{\alpha(k_i)/2\}]$.
There exists a gap in the spin excitation spectrum equal to the binding
energy of a pair and only the charge sector remains gapless.
Hence only the charge sector in the bosonized Hamiltonian is
retained and the only relevant parameters
in (\ref{eq-boson}) are $K_c$ and $v_c$. In order to make the
reference to the Cooper pairs explicit, we shall rename these parameters
$K_p$ and $v_p$.

The pair stiffness is defined as before
$D_p=(L/2)\ {\partial^2 E_0(\Phi_p)/ \partial\Phi_p^2}$ and the pair
compressibility is computed by changing the
ground state configuration by adding a Cooper pair. The
thermodynamic expression for the pair compressibility is
$\chi_p^{-1}=(m^2/ L){\partial^2 E_0(M)/ \partial m^2}$,
where $m$ is the density of Cooper pairs.
The explicit form of the derivative for a mesoscopic system is
$\chi_p^{-1}=(M^2/L)\left[E_0(M+1)+E_0(M-1)-2E_0(M) \right]$.
We can now compute $K_p$ and $v_p$ from the ground state energies.
The relations are
\begin{mathletters}
\label{eq-bosonconstatt}
\begin{eqnarray}
K_p &=& \pi m \sqrt{2 D_p\over \chi_p^{-1}}, \\
v_p &=& {1\over m}\sqrt{{D_p\chi_p^{-1}\over 2}}.
\end{eqnarray}
\end{mathletters}

The error involved in using the Cooper Pair representation is of
$O(\exp[-2 U L])$.
{}From Fig.\ \ref{fig-dcdp}, we see that down to $|U|=0.01$ there is
very good agreement of the pair stiffness in the
attractive regime with the charge stiffness found in the repulsive
regime. This justifies our use of the effective BA equations
even for such small $U$.
The general features of the pair stiffness are the same as that seen
for the charge stiffness. One sees a reduction in the pair stiffness as
the attraction is increased. Note that for small $|U|$ and density $n$,
the value of $D_c$ is always larger or equal to $D_p$.
Here the charge currents in the repulsive case are larger than in the
attractive case.
For densities close to half-filling and $U>1$, this behavior is no
longer true as $4 k_f$ scattering terms become dominant in the
repulsive regime.

%
%
\section{Effect of disorder}
\label{sec-disorder}

In the limit of weak disorder, the interaction between the
particles and disorder can be parameterized by two uncorrelated
Gaussian random fields $\eta$ and $\xi$.\cite{abrikosov} These
two fields
describe the forward and backward scattering by the impurities.
The forward scattering term can be treated
exactly\cite{apel,zawa,tgh1,tgh2} in one dimension and is found not
to contribute to the conductivity.
However, the effect of backward scattering is very
important and leads to localization in the non-interacting
limit.\cite{abrikosov}
$\xi$ and $\xi^*$ correspond to the part of the random potential
which has Fourier components close to $2k_f$.
Higher Fourier components are less effective and do not correspond to
low energy processes.
A notable exception is the $4 k_f$ term which we will have reason to
discuss in the last section.
In terms of the boson variables the impurity coupling to the
particle density is given as
\begin{equation}
 H_\Delta= \int{ dx\; \xi(x)\; e^{i\sqrt{2}\phi_{\sigma}(x)}
\cos{(\sqrt{2}\phi_{\sigma})} + h.c.},
\end{equation}
where $\xi$ is a Gaussian with
$\langle\xi(x)\xi^*(x^\prime)\rangle=\Delta_{\xi}\delta(x-x^\prime)$.
Contrary to the free case the charge and spin degrees of freedom
are no longer independent but are coupled through the random
potential.

Generally both impurity backscattering and the
interaction term $g_{1\perp}$ give rise to divergent terms in a
perturbation
calculation. Hence, a perturbative approach in the disorder
$\Delta_{\xi}$ and in the interaction $g_{1\perp}$ is used to
generate the renormalization group equations under a change of the
length scale $\alpha\rightarrow e^l \alpha$,\cite{tgh1,tgh2}
where $\alpha$ is the lattice spacing.
The equations are
\begin{mathletters}
\label{eq-rgrep}
\begin{eqnarray}
{dK_c(l)\over dl}&=&-{1\over 2}\left[{K_c^2v_c\over
v_s}\right]{\Delta}(l),\\
{dK_s(l)\over dl}&=&-{1\over 2}\left[ {\Delta}(l)+y(l)^2\right]K_s^2,\\
{dv_c(l)\over dl}&=&-{v_c^2K_c\over 2v_s}{\Delta}(l),\\
{dv_s(l)\over dl}&=&-{v_sK_s\over 2}{\Delta}(l),\\
{dy(l)\over dl}&=&\left[2-2K_s(l)\right]y(l)-{\Delta}(l),\\
{d{\Delta}(l)\over dl}&=&\left[3-K_c(l)-K_s(l)-y(l)\right]{\Delta}(l),
 \label{eq-rgrep-delta}
\end{eqnarray}
\end{mathletters}
with the dimensionless quantities $\Delta$ and $y$ defined as:
${\Delta}= (2\Delta_{\xi}\alpha / \pi v_s^2)
\left [ {v_s/ v_c}\right]^{K_c}$
and
$y=g_{1\perp}/\pi v_s$.
The development of the RG equations is limited to first
order in $\Delta_{\xi}$ and to second order in $g_{1\perp}$ and
hence the renormalizations of
$K_{r}$ and $v_{r}$ can be neglected on the right hand side
of the first three equations.

For a mesoscopic system of size $L$, the infra-red cutoff
is expected to be given by $L$ s.t.\ $L\sim e^l \alpha$, or
equivalently, $l\sim \mbox{ln} L/\alpha$. Thus we may calculate
charge and spin stiffness at finite size by using the values
of $K_c(l), K_s(l), u_c(l)$ and $c_s(l)$ and the formulas of
section \ref{sec-hub}.

The RG equations for finite disorder have to be numerically
integrated taking appropriate initial conditions at $l=0$.
Using perturbative values as initial data for the RG equations, e.g.,
$v_s=v_f$ and $K_s=1+y/2$, we see that we get an increasing value for
the spin stiffness $D_s={1\over 4\pi}v_s K_s$ with increasing repulsion.
This does not agree with the exact Bethe Ansatz solution
for the spin stiffness as shown in Fig.\ \ref{fig-ds} which clearly
shows a reducing stiffness.
Therefore, we shall use the clean Hubbard model parameters that were
computed in section \ref{sec-hub} and $g_{1\perp}=U$.
In the following sections we discuss the results of the integration
primarily for $n=0.3$.
The results for fillings corresponding to
$M= 10, 22, 42, 50, 62, 70, 82, 90$ and $96$ are qualitatively the same.

\subsection{The repulsive case}

For the repulsive case we find that $\Delta$ always flows to $\infty$
as shown in Fig.\ \ref{fig-dreprg}.
Assuming that there is no other fixed point at intermediate coupling,
this whole region can then be identified with the localized phase.
\cite{tgh1,tgh2}
The magnetic properties of the system will then depend of the
renormalized value of $y$:

(i) Increasing the repulsive interaction away from $U=0$, we find that
$y$ flows to $-\infty$ as shown in Fig.\ \ref{fig-yreprg}.
The physical state corresponds to a non-magnetic system of localized
pairs of spins which is equivalent to a charge density wave pinned by the
impurities (PCDW).\cite{tgh1,tgh2}
In Fig.\ \ref{fig-dreprg} we see that the disorder scales less
rapidly to infinity as the interaction strength is increased.
This implies that pinning of the charge density wave is harder
in the presence of a repulsive interaction. Thus we expect the
interacting charge currents to be enhanced by the disorder.
However, as shown in section \ref{sec-hub}, there is an initial
reduction in the interacting charge currents to due the presence
of the lattice.
In Fig.\ \ref{fig-dcreprg}, we show the renormalization of the charge
stiffness for a fixed filling $n=0.3$ and disorder strength for various
values of the interaction strength.
We see that there is a crossover between the non interacting and
interacting stiffnesses for {\em finite} $l_c$, or, equivalently,
finite disorder ${\Delta_{\xi}}(l_c,U)$ and we find enhancement
only for disorder values larger than ${\Delta_{\xi}}(l_c,U)$.
The crossover regions for a fixed value of the
disorder as a function of $n$ and $U$ are shown in
Fig.\ \ref{fig-repdcx}.
Previous studies of the disordered Hubbard ring\cite{kamal,ramin}
treat the Hubbard interaction in first order perturbation theory
for finite disorder and thus cannot identity this crossover.

In Fig.\ \ref{fig-yreprg} we further note that $y$ scales less
rapidly to $-\infty$ as the interaction strength is increased.
As noted earlier, the physical state corresponds
to a system of localized spins and hence larger the $U$ value lesser
the localization. We therefore find that the spin currents
are less pinned by the impurity as the repulsion is enhanced.
In Fig.\ \ref{fig-dsreprg} we again find due to the reduced pinning
effect there exists a critical disorder strength ${\Delta_{\xi}}(l_s,U)$
where there is a crossover between the non-interacting and interacting
spin currents.
$l_s$ here is the length at which the crossover occurs and
${\Delta_{\xi}}(l_s,U)$ is the value of the disorder at $l_s$ for a
given $U$.
Note that we always find $l_c < l_s$
(${\Delta_{\xi}}(l_c,U) < {\Delta_{\xi}}(l_s,U)$).
We remark that the actual value of ${\Delta_{\xi}}(l_s,U)$ is independent
of particle density $n$. However, we expect a strong dependence on the
magnetization. As we have restricted our study to the sector of zero
magnetization, we do not observe this later dependence here.

(ii) In Fig. \ref{fig-yreprg}, we see that further increase of $U$
results in $y$ flowing to $+\infty$.
Hence there is a strong repulsion of up and down spins, and the
particles start to localize as isolated spins on randomly distributed
sites. The magnetic properties have been identified earlier as
being typical of a random antiferromagnet (RAF).\cite{tgh1,tgh2}

The charge currents in the RAF phase will continue to be enhanced.
In Fig. \ref{fig-dcreprg},
we see that, e.g., the charge stiffness for $n=0.3$ and $U$ values of
$0.8$ and $1.0$ is still above the non-interacting stiffness.
Fig. \ref{fig-yreprg} clearly shows that these values of $U$ already
belong to the RAF fixed point $y^* = \infty$.

The strong repulsion of up and down spins, however, gives rise
to a drastic fall in the spin currents as seen again for
$n=0.3$ in Fig.\ \ref{fig-dsreprg}.
We thus observe {\em no} enhancements in the spin currents in the
RAF phase.
This may be used to distinguish the two different phases of the
system for repulsive $U$. In Fig.\ \ref{fig-reppcdwraf} the
region above the line represents the RAF state where no enhancement
of the spin currents is found, whereas the region below the  line
corresponds to the PCDW state.
The reader should compare Fig.\ \ref{fig-reppcdwraf} with
Fig.\ \ref{fig-repdcx}:
The maxima of each curve in Fig. \ref{fig-repdcx} corresponds to the
transition point between PCDW and RAF phases as in
Fig.\ \ref{fig-reppcdwraf}.
We note that for values of $n$ and $U$ such that the fixed point of
the system belongs to the RAF phase, already small disorder
values will localize the spins and thus lead to an enhancement of the
charge stiffness.

\subsection{The attractive case}

For negative $U$ there exists a gap in the spin excitation
spectrum and only the charge sector remains gapless.
The RG Eqs.\ (\ref{eq-rgrep}) can be reduced by taking into
account only the charge (i.e.\ pair) excitations,\cite{tgh1,tgh2}
\begin{mathletters}
\label{eq-rgatt}
\begin{eqnarray}
{dK_p(l)\over dl}&=&-{1\over 2}K_p^2{\Delta}(l),\\
{dv_p(l)\over dl}&=&-{v_p^2K_p\over 2}{\Delta}(l),\\
{d{\Delta}(l)\over dl}&=&\left[3-K_p(l)\right]{\Delta}(l),
 \label{eq-rgatt-delta}
\end{eqnarray}
\end{mathletters}
with ${\Delta}(l)=(2C_s \Delta_{\xi}\alpha / \pi v_p^2)$ and
$C_s$ a constant of order unity.
These equations have been studied previously and we find in agreement
with Ref.\ \onlinecite{tg}:
{}From Eq.\ (\ref{eq-rgatt-delta}) we see that as $K_p < 3$, the
disorder will always scale to large values thereby always leading to
localization of the particles. This implies for the second
equation that $v_p\rightarrow 0$. The stiffness therefore always
reduces in the presence of disorder. As the attraction is increased
the disorder scales faster to infinity and since in the attractive
Hubbard model the ground state contains strong charge density
fluctuations they get easily pinned to the disorder. Due to the large
reduction in the bare stiffness and an increased pinning effect to
the disorder we therefore see {\em no} enhancement in the persistent
currents as the attraction is increased.

%
%
\section{Conclusions}
\label{sec-conclusion}

We have studied in this work the behavior of charge and spin stiffness
constants of a 1D disordered mesoscopic Hubbard ring.
For the attractive Hubbard model without disorder, we find in agreement
with previous studies\cite{tg} that the charge stiffness gets reduced with
increasing $|U|$ and is always smaller than the corresponding charge
stiffness for the repulsive regime at densities not close to half-filling.
For finite disorder, we too find no enhancements in the stiffness
for finite $|U|$ over the non interacting currents.

For the repulsive Hubbard model without disorder, we again see that
the charge and spin stiffnesses get reduced as $U$ is increased.
Additionally, we observe a very strong reduction as we approach the
half-filled situation for finite $U$.
However, for finite disorder, the situation is very different from the
attractive case:
For small $U$, the inclusion of disorder drives the system into
a localized state with enhancement of both charge and spin stiffnesses.
However, for a mesoscopic system, enhancement is only observed for
large enough values of the disorder, or, equivalently, large enough
system sizes.
The physical state in this phase corresponds to a non-magnetic
system of localized pairs of spins which is equivalent to a
charge density wave pinned by the impurities (PCDW).\cite{tgh1,tgh2}
As the interaction is increased the tendency towards pinning is found
to reduce as the disorder scales less rapidly towards $\infty$.

For larger $U$, the effective interaction between
up and down spins becomes repulsive and we identify the physical state
as corresponding to isolated spins localized on randomly distributed
sites (RAF).\cite{tgh1,tgh2}
This strong repulsion between unlike spins reduces the spin current
drastically and we find {\em no} enhancements in the spin current.
However, the charge currents in the RAF phase are still enhanced
as compared to the non interacting current.

We emphasize that the transition from PCDW state to RAF state
at intermediate values of the interaction strength $U$ can be
identified by studying the different behavior of $D_s$ in the
two phases.

In this study, we have neglected Umklapp and $4 k_f$ scattering terms
both in the bosonized formulation of the Hubbard Hamiltonian and in the
RG treatment of the disorder. Thus processes close to half-filling
and $U \gg 1$ are presumably not included in our study.
In order to at least qualitatively see what happens for large $U$,
we have studied a $4$ site Hubbard ring by exact diagonalization.
Then due to the finite value of $l_c$, we need to go to large values
of the disorder strength $\Delta_\xi > 5$ in order to observe the
crossover in $D_c$.
However, we still do find a crossover even for $U=100$.

\acknowledgments
The authors would like to thank T.\ Giamarchi and B.\ S.\ Shastry
for many insights and fruitful discussions.
R.A.R.\ gratefully acknowledges financial support from the
Alexander von Humboldt foundation.


\begin{figure}
  \caption{
   The charge $D_c$ and pair $D_p$ stiffness for the Hubbard ring of
   size $L=100$, as a function of the interaction strength for different
   values of the filling $N= 2M = 10, 22, 30, 42, 50, 62, 70, 82, 90, 96$
   from bottom to top.
  }
\label{fig-dcdp}
\end{figure}

\begin{figure}
  \caption{
   The spin stiffness $D_s$  for the Hubbard ring of
   size $L=100$, as a function of the interaction strength for different
   values of the filling $N= 2M = 10, 22, 30, 42, 50, 62, 70, 82, 90, 96$
   from bottom to top.
   Note that for large $N$ and $U$, it becomes increasingly difficult to
   distinguish different $D_s$ curves due to limited numerical
   resolution.
  }
\label{fig-ds}
\end{figure}

\begin{figure}
  \caption{
   Values of the disorder $\Delta$ as a function of the system size
   $l\sim \mbox{ln} L/\alpha$ obtained by numerically
   integrating Eq.\ \protect{\ref{eq-rgrep}} for different values
   of the interaction strength
   $U = 0, 0.01, 0.05, 0.1, 0.15, 0.2, 0.3, 0.4, 0.5, 0.6, 0.8,
   1.0, 1.2, 1.4, 1.5, 1.6, 1.8, 2.0, 4., 6., 8., 10., 15$ and $20$
   (small $U$ is represented by small dashes) for a fixed density
   $n=0.3$.
   The initial value for the disorder is fixed to
   $\Delta/v_f = 5 \times 10^{-4}$.
   The disorder scales less rapidly to infinity as
   the interaction strength is increased.
  }
\label{fig-dreprg}
\end{figure}

\begin{figure}
  \caption{
   Values of $y$ as a function of the system size
   $l\sim \mbox{ln} L/\alpha$.
   Density and initial disorder strength are as in
   Fig.\ \protect{\ref{fig-dreprg}},
   $U$ varies from $0$ to $20$.
   Increasing $U$ is indicated by increasing dash length.
   $y$ scales less rapidly to $-\infty$ as
   the interaction strength is increased from $0$ to $0.6$.
   Starting at $U=0.8$, $y$ then scales to $+\infty$.
  }
\label{fig-yreprg}
\end{figure}

\begin{figure}
  \caption{
   Values of the charge stiffness $D_c$ as a function of
   the system size $l\sim \mbox{ln} L/\alpha$.
   Density and initial disorder strength are as in
   Fig.\ \protect{\ref{fig-dreprg}}.
   For clarity, $U$ is only shown for values equal to
   $0.0, 0.1, 0.2, 0.4, 0.6, 0.8$ and $1$.
   Increasing $U$ is indicated by increasing dash length.
   There exists a crossover region beyond
   which enhancement of the currents is observed.
   Note that the lines for $U=0.8$ and $U=1$ correspond to
   the RAF phase and are well above the $U= 0$ value.
  }
\label{fig-dcreprg}
\end{figure}

\begin{figure}
 \caption{
  We plot the disorder strength
  $\Delta_\xi(l_c)= (\pi v_s^2 \Delta/2 \alpha) [v_c/v_s]^{K_c}$
  at which the enhancement of interacting over non interacting
  $D_c$ occurs as function of interaction strength $U$.
  Different curves correspond to different fillings, i.e.,
  $N= 2M = 10, 22, 30, 42, 50, 62, 70, 82, 90, 96$
  from bottom to top.
  Note that we have only limited resolution in $U$ corresponding
  to the possible values of the interaction strength as in Fig.
  \protect{\ref{fig-dreprg}}.
 }
\label{fig-repdcx}
\end{figure}

\begin{figure}
  \caption{
   Values of the spin stiffness $D_s$ as a function of the
   system size $l\sim \mbox{ln} L/\alpha$.
   Density and initial disorder strength are as in
   Fig.\ \protect{\ref{fig-dreprg}},
   $U$ varies from $0$ to $20$.
   Increasing $U$ is indicated by increasing dash length.
   There exists a crossover region beyond
   which enhancement of the currents is observed for
   $0 \leq U \leq 0.6$.
   After the PCDW-RAF transition between $0.6 < U < 0.8$,
   $D_s$ rapidly scales to zero.
  }
\label{fig-dsreprg}
\end{figure}

\begin{figure}
  \caption{
   Using the qualitatively different behavior of $D_s$, we identity
   the boundary between the PCDW and RAF phases.
   The data points represent the largest values of $U$ for a given
   filling at which we still see a crossover in $D_s$ and are thus
   part of the PCDW phase.
   Note that we have only a limited resolution in $U$ corresponding to
   the possible values of the interaction strength as in
   Fig.\ \protect{\ref{fig-dreprg}}.
  }
\label{fig-reppcdwraf}
\end{figure}

\end{document}